\documentclass{article}
\usepackage{amssymb}
\usepackage{graphicx}
\usepackage{cite}
\begin{document}
\title{On the (2+1)-dimensional Dirac equation in a constant magnetic field
with a minimal length uncertainty}
\author{P. Pedram\thanks{Email:
p.pedram@srbiau.ac.ir}, M. Amirfakhrian, and H. Shababi
\\  {\small Department of Physics, Science and Research Branch,
Islamic Azad University, Tehran, Iran}}

\maketitle \baselineskip 24pt

\begin{abstract}
We exactly solve the (2+1)-dimensional Dirac equation in a constant
magnetic field in the presence of a minimal length. Using a proper
ansatz for the wave function, we transform the Dirac Hamiltonian
into two 2-dimensional non-relativistic harmonic oscillator and
obtain the solutions without directly solving the corresponding
differential equations which are presented by Menculini et al.
[Phys. Rev. D 87, 065017 (2013)]. We also show that Menculini et al.
solution is a subset of the general solution which is related to the
even quantum numbers.

\vspace{5mm} {\it PACS numbers:} 04.60.Bc

\vspace{.5cm} {\it Keywords}: Generalized Uncertainty Principle;
Minimal length; Dirac equation
\end{abstract}
\maketitle

\section{Introduction}
At high energy limit, close to the Planck scale where the
corresponding Schwarzschild radius becomes comparable with the
Compton wavelength and both tend to the Planck length, the effects
of gravity become so important that would result in discreetness of
the spacetime. In this case, different approaches to quantum gravity
such as string theory
\cite{Veneziano,Amati89,Amati87,Gross,Konishi}, noncommutative
geometry \cite{Capozziello}, and loop quantum gravity \cite{Garay}
predict the existence of a minimal measurable length. These theories
argue that near the Planck scale, the Heisenberg Uncertainty
Principle should be replaced by the so-called Generalized
Uncertainty Principle (GUP).

Recently, various studies about the effects of the minimal length
have been done in the literature such as hydrogen atom spectrum
\cite{Brau,Benczik,Stetsko.Tkachuk,Stetsko}, Lamb shift
\cite{Benczik,Stetsko.Tkachuk,Nouicer2006}, harmonic oscillator
\cite{Quesne2003,Quesne2004,Quesne2005,Chang,Kempf95,Kempf97},
gravitational quantum well \cite{Brau.Buisseret},  Casimir effect
\cite{Nouicer,Harbach.Hossenfelder}, particles scattering
\cite{Hossenfelder,Stetsko.Tkachuk2007,Pikovski}, resolution of wave
function singularities for strongly attractive potentials
\cite{Bouaziz2007}, and the classical limit of the minimal length
uncertainty \cite{Benczik2002}. Also, the effects of the GUP on the
electromagnetic field is discussed in Ref.~\cite{Camacho}.

The effects of the minimal length and maximal momentum on
relativistic and non-relativistic wave equations have been
investigated in Refs. \cite{Pedram2,Das2,Das1,Das3}. The problems of
Lorentz violation, Dirac particle in a box,  Dirac oscillator
\cite{Pedram2}, potential step, and potential barrier \cite{Das2}
are studied in the presence of the minimal length uncertainty. Using
a recent proposal by Ali, Das and Vagenas that implies a minimal
length uncertainty and a maximal momentum \cite{Das3}, the problems
of superconductivity and the quantum Hall effect are discussed in
Ref.~\cite{Das1}. Moreover, the solutions of the 3-dimensional Dirac
oscillator in the presence of the minimal length are presented
Ref.~\cite{Quesne2005}.

The solutions of the Dirac equation in the presence of a homogeneous
magnetic field have many applications in condensed matter physics.
Recently, for the 2-dimensional electron systems, \emph{i.e.}
non-relativistic Landau levels, the expected ring like internal
structure of the wave functions is detected for the first time using
scanning tunneling spectroscopy \cite{Hashimoto}. Also, the massless
(2+1)-dimensional Dirac equation is used to describe the motion of
electrons in graphene \cite{A.K.Geim} and to find an upper bound on
the fundamental minimal length $\hbar\sqrt{\beta}$ using
experimental measurements of the relativistic Landau levels in
graphene \cite{Jiang}. Notice that, if we take $\beta$ as a
non-universal parameter, this upper bound could be varied from one
experiment to another. For instance, in the Lamb shift an upper
bound for the minimal length is obtained
$\hbar\sqrt{\beta}<10^{-17}m$ which is of the order of the
electroweak scale \cite{Stetsko,Benczik2002}. But, for the ultracold
neutron energy levels in a gravitational quantum well
\cite{Brau.Buisseret,Pedram}, the upper bound is $\hbar\sqrt{\beta}<
2.41\times 10^{-9}m$ which agrees well with the results that
obtained for the massless (2+1)-dimensional Dirac equation
\cite{L.Menculeni}.

In this paper, we study the problem of the (2+1)-dimensional Dirac
equation in the presence of a constant magnetic field and a minimal
length uncertainty. This problem is recently investigated in
Refs.~\cite{L.Menculeni,Moniruzzaman} by directly solving the
corresponding differential equations in momentum space. However, as
we shall show, using a proper ansatz for the momentum space wave
function, the Dirac Hamiltonian can be cast into two 2-dimensional
non-relativistic harmonic oscillators which is exactly solvable in
terms of Jacobi polynomials \cite{Chang}. Then, without directly
solving the differential equations, we find exact energy eigenvalues
and eigenfunctions and show that Menculini \emph{et al.} results
correspond to the even quantum numbers as a part of the general
solution.

\section{The generalized uncertainty principle}
Let us consider the generalized uncertainty principle in the form
\cite{Kempf95}
\begin{equation}\label{1}
\Delta X_i \Delta P_j\geq\frac{\hbar }{2}{\delta }_{ij}\left(1+\beta
\left(\left(\Delta P\right)^{2}+ {\left\langle P\right\rangle
}^2\right)\right),
\end{equation}
where $\beta$ is the GUP parameter. The above inequality relation
leads to the existence of a minimal measurable length $(\Delta
X)_{min}=\hbar \sqrt{\beta}$ which is of the order of the Planck
length $\ell _{\mathrm{Pl}}=\sqrt{\frac{G\hbar}{c^{3}}}\approx
10^{-35}$m \cite{Kempf95,Kempf97,Kempf94}. This uncertainty relation
leads to the following deformed commutation relation, namely
\begin{equation}\label{2}
\left [X_{i},P_{j}\right]=i\hbar\delta_{ij}\left(1+\beta
P^{2}\right),
\end{equation}
where $P^2=\sum_i P_i^2$. It is straightforward to check that when
$\beta = 0$, the well-known commutation relation in ordinary quantum
mechanics is recovered. In momentum space representation, we have
\begin{eqnarray}\label{3}
P_{i}\psi(p)&=&p_{i}\psi(p),\\ \label{4}
X_{i}\psi(p)&=&i\hbar(1+\beta {P}^ {2})\partial_{p_{i}}\psi(p).
\end{eqnarray}
Now, using Eq.~(\ref{4}), the commutation relations for position
operators reads
\begin{eqnarray}\label{5}
[X_i,X_j ]=2i\hbar\beta  (P_i X_j-P_j X_i ),
\end{eqnarray}
as a noncommutative generalization of the position space. Note that,
the rotational symmetry is not broken by the commutation relations
(\ref{2}) and (\ref{5}). In fact, we can still express the
generators of rotations in terms of the position and momentum
operators as
\begin{eqnarray}\label{8}
L_{ij}={\frac{1}{1+\beta \vec{P}^2 } (X_i P_j-X_j P_i)},
\end{eqnarray}
where in the limit $\beta \rightarrow 0$, we obtain the ordinary
definition of orbital angular momentum.

\section{Dirac equation in the GUP framework}
The Dirac Hamiltonian in the presence of a homogeneous magnetic
field $  \vec{B}=(0,0,B_0)$  takes the form
\begin{eqnarray}\label{10}
H=c{\vec{\sigma} }.\left(\vec{P}+\frac{e}{c}{\vec{A}}\right)+{\sigma
}_z Mc^2,
\end{eqnarray}
where, $\vec{\sigma}$  and  $\vec{A}$ denote the Pauli matrices
 and the vector potential in the symmetric
gauge, respectively
\begin{eqnarray}\label{11}
A_x=-\frac{1}{2}B_0Y,   \hspace{2cm} A_y=\frac{1}{2}B_0X.
\end{eqnarray}
The eigenvalue problem is
\begin{eqnarray}\label{12}
H\psi(p) =E\psi(p) , \hspace{1cm}      \psi(p) =\left(
\begin{array}{c}
{\psi }^{(1)}(p)\\
{\psi }^{(2)}(p) \end{array} \right).
\end{eqnarray}
If we define $P_{\pm }$ as
\begin{eqnarray}\label{13}
P_{\pm }=P_x\pm iP_y=\left(P_x+\frac{e}{c}A_x\right)\pm
i\left(P_y+\frac{e}{c}A_y\right),
\end{eqnarray}
the eigenvalue equation can be written as
\begin{eqnarray}\label{14}
H\psi(p) =\left( \begin{array}{cc}
Mc^2 & cP_- \\
cP_+ & -Mc^2 \end{array} \right)\left( \begin{array}{c}
{\psi }^{(1)} (p)\\
{\psi }^{(2)}(p) \end{array} \right)=E\left( \begin{array}{c}
{\psi }^{(1)}(p) \\
{\psi }^{(2)}(p) \end{array} \right).
\end{eqnarray}
Thus, we have
\begin{equation}\label{15}
P_- {\psi }^{(2)}(p)=\epsilon_- {\psi }^{(1)}(p), \hspace{1cm}  P_+
{\psi }^{(1)}(p)=\epsilon_+ {\psi }^{(2)}(p),
\end{equation}
where $\epsilon_{\pm }=(E\pm Mc^2)/c$. By separating the components
of Eq.~(\ref{14}) and using Eq.~(\ref{15}) we find
\begin{eqnarray}\label{16}
H^{\left(1\right)}{\psi }^{(1)}(p)&=&P_- P_+ {\psi
}^{(1)}(p)=\epsilon^2{\psi }^{(1)}(p),\\
\label{17}
 H^{\left(2\right)}{\psi }^{(2)}(p)&=&\ P_+ P_- {\psi }^{(2)}(p)=\epsilon^2{\psi
 }^{(2)}(p),
\end{eqnarray}
where $\epsilon^2=\epsilon_+ \epsilon_- =(E^2-M^2c^4)/c^2$.

Before proceed further, let us consider the problem of the
non-relativistic harmonic oscillator in two dimensions which is
exactly solvable \cite{Chang}. Its Hamiltonian is given by
\begin{equation}
H=\frac{1}{2\mu}P^2+\frac{1}{2}\mu\omega^2{(X^2+Y^2)},
\end{equation}
where the eigenvalue equation is given by $H\psi(p) =E\psi(p)$.
Since this Hamiltonian is rotationally symmetric, the energy
eigenfunction can be written as a product of a radial wave function
and spherical harmonics. So, we have
\begin{equation}\label{19}
\psi(p)=\frac{1}{\sqrt{2\pi }}e^{im\phi}R(p).
\end{equation}
Here, $m$ is the quantum number associated to the operator $L_{z}$
and $R(p)$ is the radial part of the wave function where in two
dimensions is given by \cite{Chang}
\begin{equation}\label{20}
R^{\,a}_{n,m}(p)={\sqrt{\frac{2\beta(2{n}'+a+|m|+1){n}'! {\Gamma}
({n}'+a+|m|+1)}{{\Gamma} ({n}'+a+1){\Gamma} ({n}'+|m|+1)}}}
{\beta}^{|m|/2} {(1+\beta p^2)}^{-(a+2+|m|)/2}
p^{|m|}P_{n'}^{(a,|m|)} (z),
\end{equation}
where $z=\frac{\beta p^2-1}{\beta p^2+1}$, ${n}'=(n-|m|)/2$, $n$ is
principal quantum numbers, $a=\sqrt{1+m^2+k^{-4}}$,
$k=\sqrt{\mu\hbar\omega\beta}$, and $P_{n}^{\left(a,|m|\right)} (z)$
is the Jacobi polynomial. Also, its energy spectrum reads
\begin{equation}\label{49}
E_{n,m}=\hbar \omega \left[(n+1)\sqrt{1+\beta^2(m^2+1)\mu^2 \hbar^2
\omega^2}+\frac{\mu\hbar\omega\beta}{2}\left((n+1)^2+m^2+1\right)\right].
\end{equation}

\section{The exact solution of Dirac equation with algebraic method }
Using Eqs.~(\ref{11}), (\ref{13}) and (\ref{16}),  the Hamiltonian
for the first component of the spinor is given by
\begin{eqnarray}
H^{\left(1\right)}&=&P_- P_+=P^2_x+P^2_y+{\alpha }^2
(X^2+Y^2)+2\alpha (XP_y- YP_x)+i{\alpha }^2 (XY-YX)-i\alpha
[X,P_x]-i \alpha [Y, P_y],\nonumber\\
 &=&(1+2\alpha\beta\hbar)P^2 +{\alpha}^2
{(X^2+Y^2)}+2\alpha (1+\beta P^2)L_z+2\beta {\alpha }^2\hbar
(1+\beta P^2)L_z+2\alpha\hbar,\label{47}
\end{eqnarray}
where $ \alpha= eB_0/(2c)$. Using the ansatz
\begin{equation}
{\psi }^{(1)} \left(p\right)=\frac{1}{\sqrt{2\pi
}}e^{im\phi}R^{\left(1\right)}(p),
\end{equation}
that satisfies $L_{z}{\psi }^{(1)}(p)= m\hbar{\psi }^{(1)}(p)$ we
obtain
\begin{equation}\label{48}
H^{\left(1\right)}{\psi }^{(1)}(p) =\left\{\left[1+2\alpha \beta
\hbar +2\alpha \beta \hbar m (1+\alpha \beta
\hbar)\right]P^2+{\alpha }^2{(X^2+Y^2)}+2\alpha \hbar m(1+\alpha
\beta \hbar )+2\alpha \hbar\right\}{\psi }^{(1)}(p).
\end{equation}
Now, this equation is similar to the Hamiltonian equation of the
harmonic oscillator, namely
$H^{(1)}=\frac{1}{2\mu_{1}}P^2+\frac{1}{2}\mu_{1}\omega_{1}^2(X^2+Y^2)+c_1$,
where $\mu_{1}=\big[2+4\alpha\beta\hbar+4\alpha\beta\hbar
m(1+\alpha\beta\hbar)\big]^{-1}$,
$\omega_{1}^2=2\alpha^2\big[2+4\alpha\beta\hbar+4\alpha\beta\hbar
m(1+\alpha\beta\hbar)\big]$, and $c_1=2\alpha \hbar m(1+\alpha \beta
\hbar )+2\alpha \hbar$. So, the normalized radial part of energy
eigenfunction for the first component is given by
\begin{equation}\label{52}
R^{\left(1\right)}(p)=R^{\,a_{1}}_{n,m} (p),
\end{equation}
where $a_{1}=\sqrt{1+m^2+k_1^{-4}}$ and
$k_{1}=\sqrt{\mu_{1}\omega_{1}\hbar\beta}$. Also, the eigenvalues of
the Hamiltonian $H^{(1)}$ are easily obtained
\begin{equation}\label{50}
{\epsilon}^{2\left(1\right)}_{n,m}=2\hbar\alpha(m+n+2)+\beta\hbar^{2}\alpha^{2}(m+n)^2+4\beta\hbar^2\alpha^2(m+n+1),
\end{equation}
and the energy spectrum reads
\begin{equation}\label{51}
{E}_{n,m}=\pm\sqrt{M^2c^4+2\hbar\alpha
c^2(m+n+2)[1+\frac{\hbar\alpha\beta}{2}(m+n+2)]}.
\end{equation}

For the second component of the spinor, the  Hamiltonian is
\begin{eqnarray}\nonumber\label{53}
H^{\left(2\right)}&=&P_+P_- =P^2_x+P^2_y+{\alpha }^2(X^2+Y^2)+2\alpha (XP_y-YP_x)-i{\alpha }^2(XY-YX)+i\alpha [X,P_x]+i\alpha [Y,\ P_y],\\
&=&(1-2\alpha\beta\hbar)P^2+\alpha^2 {(X^2+Y^2)}+2\alpha (1+\beta
P^2)L_z-2\beta
 {\alpha }^2\hbar (1+\beta P^2)L_z-2\alpha \hbar.
\end{eqnarray}
Using the ansatz
\begin{equation}\label{35}
{\psi }^{(2)} \left(p\right)=\frac{1}{\sqrt{2\pi
}}e^{im'\phi}R^{\left(2\right)}(p),
\end{equation}
and $L_{z}{\psi }^{(2)}(p)= m'\hbar{\psi }^{(2)}(p)$ we obtain
\begin{equation}\label{54}
H^{(2)}{\psi }^{(2)}(p)=\{[1-2\alpha \beta \hbar +2\alpha \beta
(1-\alpha \beta \hbar ){m}'\hbar]P^2+{\alpha }^2{(X^2+Y^2)}+2\alpha
(1-\alpha \beta \hbar ){m}'\hbar -2\alpha \hbar \}{\psi^{(2)}}(p).
\end{equation}
This Hamiltonian is again similar to the harmonic oscillator
Hamiltonian
$H^{(2)}=\frac{1}{2\mu_{2}}P^2+\frac{1}{2}\mu_{2}\omega_{2}^2{(X^2+Y^2)}+c_2$,
where
$\mu_{2}=[2-4\alpha\beta\hbar+4\alpha\beta\hbar{m}'(1-\alpha\beta\hbar)]^{-1}$,
$\omega_{2}^2=2\alpha^2\big[2-4\alpha\beta\hbar+4\alpha\beta\hbar{m}'(1-\alpha\beta\hbar)\big]$,
and $c_2=2\alpha (1-\alpha \beta \hbar ){m}'\hbar -2\alpha \hbar$.
Therefore, the normalized radial part of the solution for the
second component is given by
\begin{equation}
R^{\left(2\right)}(p)=R^{\,a_{2}}_{n,{m}'}(p),
\end{equation}
where $a_{2}=\sqrt{1+{m}'^2+k_2^{-4}}$, and
$k_{2}=\sqrt{\mu_{2}\omega_{2}\hbar\beta}$. Also, the eigenvalues of
Eq.~(\ref{54}) read
\begin{equation}\label{55}
{\epsilon}^{2\left(2\right)}_{n,{m}'}=\beta\hbar^2\alpha^2({m}'+n)^2+2\hbar\alpha({m}'+n),
\end{equation}
and the energy spectrum can be written as
\begin{equation}\label{56}
{E}_{n,{m}'}=\pm\sqrt{M^2c^4+2\hbar\alpha
c^2({m}'+n)[1+\frac{\hbar\alpha\beta}{2}({m}'+n)]}.
\end{equation}
Note that, since ${\epsilon}^{2\left(1\right)}_{n,{m}}$ should be
equal to ${\epsilon}^{2\left(2\right)}_{n,{{m}'}}$, using
Eqs.~(\ref{50}) and (\ref{55}), we obtain ${m}'= m+2$.

For a particular case, suppose that the quantum numbers take even
values, i.e. $(n,m,{m}')\rightarrow(2n,2m,2{m}')$. For the first
component we obtain
\begin{equation}\label{59}
{\epsilon}^{2\left(1\right)}_{n,{m}}=4\hbar\alpha(m+n+1)+4\beta\hbar^{2}\alpha^{2}(m+n)^2+8\beta\hbar^2\alpha^2(m+n)+4\beta\hbar^{2}\alpha^{2},
\end{equation}
and
\begin{equation}\label{60}
{E}_{n,{m}}=\pm\sqrt{M^2c^4+4\hbar\alpha
c^2(m+n+1)[1+\hbar\alpha\beta(m+n+1)]},
\end{equation}
in agreement with Ref.~\cite{L.Menculeni}. Also, for the second
component we have
\begin{equation}\label{62}
{\epsilon}^{2\left(2\right)}_{n,{{m}'}}=4\beta\hbar^2\alpha^2({m}'+n)^2+4\hbar\alpha({m}'+n),
\end{equation}
and
\begin{equation}\label{63}
{E}_{n,{{m}'}}=\pm\sqrt{M^2c^4+4\hbar\alpha
c^2({m}'+n)[1+\hbar\alpha\beta({m}'+n)]}.
\end{equation}
Now, the condition
${\epsilon}^{2\left(1\right)}_{n,{m}}={\epsilon}^{2\left(2\right)}_{n,{{m}'}}$
implies ${m}'=m+1$ that agrees with Ref.~\cite{L.Menculeni}. These
results show that the set of solutions obtained by Menculini
\emph{et al.} is a subset of general solutions presented in this
section which corresponds to the even quantum numbers.

For the massless (2+1)-dimensional Dirac equation that describes the
motion of electrons in new materials such as graphene, the
Hamiltonian is given by
\begin{equation}\label{64}
H=v_{F}{\vec{\sigma} }\cdot\left(\vec{P}+\frac{e}{c}\vec{A}\right).
\end{equation}
Here, $v_{F}$ is the Fermi velocity which for electrons in graphene
this velocity is more than the speed of light. For this Hamiltonian,
the energy eigenvalues read
\begin{equation}\label{65}
E_{n,m}=v_{F}\sqrt{2\hbar\alpha\left(m+n\right)\left[1+\frac{\hbar\alpha\beta}{2}\left(m+n\right)\right]}.
\end{equation}
Using this energy spectrum an upper bound on the fundamental minimal
length can be estimated by comparison with the experimental results
of the relativistic Landau levels in graphene \cite{Jiang}. For
example, for $B_{0}=18\,T$ and $v_{F}=(1.12\pm0.02)\times 10^6
\,$m/s, the first experimental exited level of the graphene Landau
spectrum in the absence of GUP is $E=(172\pm3)$meV \cite{Jiang}. So,
by setting $n=1$ and $m=0$ in Eq.~(\ref{65}), we obtain
\begin{eqnarray}
E^{\left(\beta\right)}_{1,0}=v_{F}\sqrt{2\hbar\alpha\left(1+\frac{\hbar\alpha\beta}{2}\right)}
=E^{\left(\beta=0\right)}_{1,0}\sqrt{1+\frac{\hbar\alpha\beta}{2}}.\label{66}
\end{eqnarray}
Now, since $\Delta
E=E^{\left(\beta\right)}_{1,0}-E^{\left(\beta=0\right)}_{1,0}<
6\,$meV, we have
\begin{equation}\label{68}
\Delta
E=E^{\left(\beta=0\right)}_{1,0}\left(\sqrt{1+\frac{\hbar\alpha\beta}{2}}-1\right)<
6\,\mbox{meV}.
\end{equation}
If we define $\delta=\beta\hbar\alpha/2=\alpha(\hbar\sqrt{\beta})^2
/(2\hbar)$, we obtain $\delta< 0.07$. Therefore, the upper bound of
the minimal length is found as
\begin{equation}\label{67}
(\Delta X)_{min}=\hbar\sqrt{\beta}<3.25\,\,\mbox{nm},
\end{equation}
which agrees with the ultracold neutron energy levels in a
gravitational field in the presence of a minimal length
\cite{Brau.Buisseret} and a minimal length and a maximal momentum
\cite{Pedram}. Note that the intermediate length scales obtained in
Ref.~\cite{Das2} are $\ell_{\mathrm{inter}}\sim
10^{10}\ell_{\mathrm{Pl}}$, $10^{18}\ell_{\mathrm{Pl}}$ and
$10^{25}\ell_{\mathrm{Pl}}$ for the potential barrier, Lamb shift,
and Landau levels, respectively. Also, based on current experiments
in superconductivity and muon experiments, the intermediate length
scales are given by $\ell_{\mathrm{inter}}\sim
10^{17}\ell_{\mathrm{Pl}}$ and $10^{8}\ell_{\mathrm{Pl}}$,
respectively \cite{Das1}. Thus, our result (\ref{67}) also agrees
with the upper bound predicted by the Landau levels \cite{Das2}.
Although our result is far weaker than the upper bound predicted by
electroweak measurements, but it is not incompatible with it.

\section{Conclusions}
In this paper, we exactly solved the (2+1)-dimensional Dirac
equation in a constant magnetic field in the framework of the
generalized uncertainty principle. Using proper ansatzs for the
momentum space wave functions, we transformed the Hamiltonian for
each component of spinor into  non-relativistic harmonic oscillator
Hamiltonians. Then, the solutions are obtained without directly
solving the GUP-corrected Dirac equation. We also showed that
Menculini \emph{et al.} solutions correspond to the even quantum
numbers as a subset of the general solution. For the massless case,
we used the experimental results to find an upper bound for the
deformation parameter which agreed with the ultracold neutron energy
levels in a gravitational quantum well.

\end{document}